\documentclass[twocolumn,aip,rsi]{revtex4}

\usepackage{graphicx}% Include figure files
\usepackage{dcolumn}% Align table columns on decimal point
\usepackage{bm}% bold math

\usepackage{xcolor}

\usepackage{booktabs}
\usepackage{chemformula}

\usepackage{setspace}
\usepackage{threeparttable}

\begin{document}

\preprint{APS/123-QED}

\title{Unexpected behaviour of the crystal growth velocity at the hypercooling limit}

\author{P. Fopp}
\email{patrick.fopp@dlr.de}
%\altaffiliation{Institut f\"ur Materialphysik im Weltraum, Deutsches Zentrum f\"ur Luft- und Raumfahrt (DLR), 51170 K\"oln, Germany}
\author{W. Hornfeck}
%\email{hornfeck@fzu.cz}
\altaffiliation{Institute of Physics of the Academy of Sciences of the Czech Republic, Na Slovance 2, 18221 Prague, Czech Republic}
\author{F.~Kargl}
%\email{florian.kargl@dlr.de}
%\altaffiliation{Institut f\"ur Materialphysik im Weltraum, Deutsches Zentrum f\"ur Luft- und Raumfahrt (DLR), 51170 K\"oln, Germany}
\altaffiliation[Also at: ]{Foundry Institute, Faculty of Georesources and Materials Engineering, RWTH Aachen University, 52062 Aachen, Germany}
\author{M. Kolbe}
%\email{matthias.kolbe@dlr.de}
%\altaffiliation{Institut f\"ur Materialphysik im Weltraum, Deutsches Zentrum f\"ur Luft- und Raumfahrt (DLR), 51170 K\"oln, Germany}
\author{R. Kobold}
\email{raphael.kobold@dlr.de}
%\altaffiliation{Institut f\"ur Materialphysik im Weltraum, Deutsches Zentrum f\"ur Luft- und Raumfahrt (DLR), 51170 K\"oln, Germany}
\altaffiliation[Also at: ]{Programmatik Raumfahrtforschung und -technologie, Deutsches Zentrum f\"ur Luft- und Raumfahrt (DLR), 51170 K\"oln, Germany}
\affiliation{Institut f\"ur Materialphysik im Weltraum, Deutsches Zentrum f\"ur Luft- und Raumfahrt (DLR), 51170 K\"oln, Germany}

\date{\today}

\begin{abstract}
The crystal growth velocity is one thermodynamic parameter of solidification experiments of undercooled melts under non-equilibrium conditions, which is directly accessible to observation. We applied the electrostatic levitation technique in order to study the crystal growth velocity $v$ as a function of the undercooling $\Delta T$ for the intermetallic, congruently melting binary alloy NiTi and the glass forming alloy Cu--Zr, as well as for the Zr-based ternary alloys (Cu$_{\mathrm{x}}$Ni$_{\mathrm{1-x}}$)Zr ($x= 0.7, 0.6$) and the Ni-based ternary alloy Ni(Zr$_{\mathrm{x}}$Ti$_{\mathrm{1-x}}) (x= 0.5)$.
All investigated systems within this work, except the eutectics \ch{Cu56Zr44} and \ch{Cu46Zr54}, exceeded the hypercooling limit $\Delta T_{\mathrm{hyp}}$ and, remarkably, every $v(\Delta T)$ relation changed significantly at $\Delta T_{\mathrm{hyp}}$.
Our results for glass forming CuZr indicate that the influence of the diffusion coefficient $D(T)$ on $v(\Delta T)$ at high undercoolings, as claimed in literature, cannot be the sole reason for the existence of a maximum in the $v(\Delta T)$ behaviour. These observations could make a valuable contribution concerning an extension of growth theories to undercooling temperatures $\Delta T > \Delta T_{\mathrm{hyp}}$. Nevertheless, our finding has direct consequences to various disciplines, as our earth and all living beings are examples for non-equilibrium systems.
The scatter of our velocity data is at least two orders of magnitude smaller than measurements performed by former works due to our experimental setup, which allowed precise contactless triggering at a specific undercooling, and our analysis method, which considered the respective solidification morphologies.
\end{abstract}

\maketitle

\section{Introduction}
\begin{figure}[tb]
	\begin{center}
		\includegraphics[width=8.6cm]{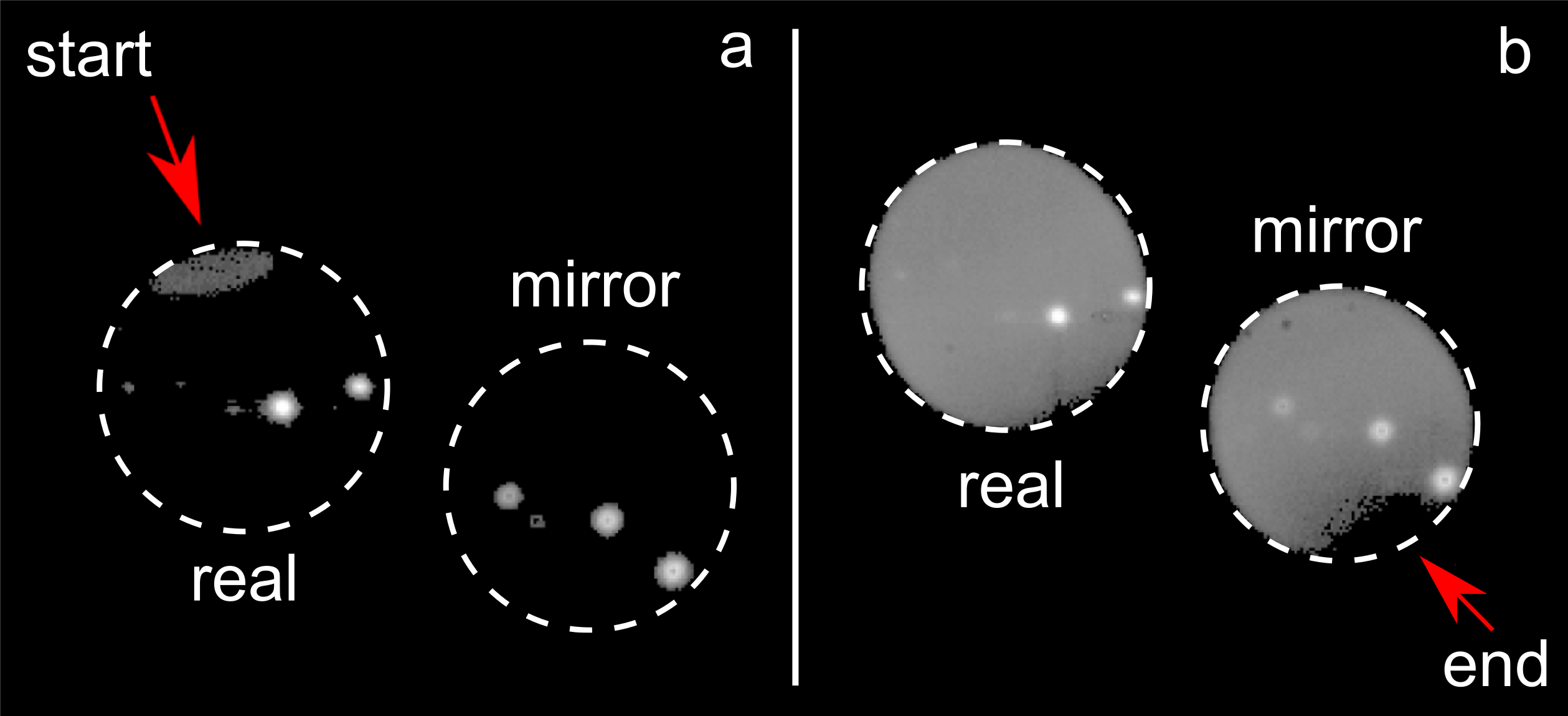}
	\end{center}
	\caption{\label{figproccham} \small{Illustration of two HSC frames, which show a) the starting point of the solidification on the frontside of the sample and b) the moment shortly before the end of solidification on the backside. A mirror installed in the levitation furnace allows to observe $\approx$ 98$\,\%$ of the sample surface, which makes it possible to accurately determine the time frame between start and end of the solidification, even for rotating samples.}}
	\label{HSC_frames}
\end{figure}

Levitation techniques offer the opportunity to investigate the solidification behaviour of metallic melts within a broad undercooling regime \cite{PaIsLeHoMoBrRhOk14,BrLoSchHoSchEg06,He14,HeKoKl18}. The range of undercooling of the freely suspended melt droplets strongly depends on the cooling rate, which in turn depends on the properties of the system, e.g. the specific heat capacity $c_{\mathrm{p}}$. Electrostatically levitated single component metallic melts usually exhibit cooling rates of about $100-200\,\mathrm{Ks^{\text{-}1}}$, whereas binary and multicomponent systems exhibit cooling rates which are $<100\,\mathrm{Ks^{\text{-}1}}$. In general, the higher the cooling rate, the higher undercoolings can be achieved, if the system is free of contaminants, which could act as easy nucleation sites. However, despite of their small cooling rates, some metallic melts, especially multicomponent systems, can reach such high undercoolings that the temperature of the melt passes the glass transition temperature $T_{\mathrm{g}}$. At $T_{\mathrm{g}}$ the structure of the melt is preserved and no change in the enthalpy $H_{\mathrm{f}}$ can be observed, i.e. the enthalpy of fusion $\Delta H_{\mathrm{f}} = 0$. In this case, the as-solidified sample exhibits an amorphous structure. For systems where the solidification is accompanied by a change in the enthalpy (i.e. $\Delta H_{\mathrm{f}} > 0$), the solidification process is visible to the unaided eye seen as sudden change in brightness of the levitating sample. With a high-speed camera the onset and end of solidification can be clearly pinpointed on the sample surface (Fig. \ref{HSC_frames}), which makes it possible to determine the (surface) solidification velocity as a function of the undercooling $v(\Delta T)$. If a broad undercooling regime is accessible, one can investigate e.g. the influence of temperature-dependent thermodynamic parameters on the $v(\Delta T)$ curve \cite{WaWaMaBiVoHeWaXuTiLi11,WaHeLi14}.

CuZr is an alloy which can form a metallic glass and has recently been investigated with respect to its dendrite growth dynamics \cite{WaHeLi14,KoKuWaHoKoHe17}. Most notably, Tang and Harrowell reported an anomalously slow crystal growth of CuZr at high undercooling temperatures \cite{TaHa13}. In \cite{WaHeLi14} it is claimed that at high undercooling temperatures the diffusion coefficient $D(T)$ becomes dominant in comparison to the driving force for solidification leading to a decrease in the slope, which ultimately becomes negative at even higher undercoolings, of the $v(\Delta T)$ curve. This leads to a maximum in $v(\Delta T)$. If the temperature of the undercooled melt is approaching $T_{\mathrm{g}}$, the characteristics of the dendrite growth kinetics changes. Similar results have already been presented for non-metallic glass formers, e.g. the organic tris-(naphthylbenzene) \cite{EdHaYu08}.

For most systems the glass transition temperature is far away from the undercoolings which can be achieved. For binary systems it is possible to reach $T_{\mathrm{g}}$, although still very difficult, especially with levitation techniques due to the limited cooling rate. The alloys investigated within this work tend to undercool up to $300-400\,\mathrm{K}$ below their equilibrium melting temperature $T_{\mathrm{m}}$. This temperature range is still far away from $T_{\mathrm{g}}$, but is sufficient for the melt to pass through the so-called \textit{hypercooling limit}, which is thermodynamically defined as
\begin{equation}
\Delta T_{\mathrm{hyp}} = \frac{\Delta H_{\mathrm{f}}}{c_{\mathrm{p}}},
\label{eq:Thyp}
\end{equation}

where complete isenthalpic solidification is achieved. For $\Delta T < \Delta T_{\mathrm{hyp}}$ the heat of fusion $\Delta H_{\mathrm{f}}$, which is released during the solidification process, is sufficient to raise the temperature of the sample to $T_{\mathrm{m}}$. In this case a certain amount (depending on the undercooling) of the material crystallizes in the time frame of the recalescence, whereas the residual melt solidifies under equilibrium conditions. In the case of $\Delta T > \Delta T_{\mathrm{hyp}}$, $\Delta H_{\mathrm{f}}$ is not sufficient to raise the temperature of the sample to $T_{\mathrm{m}}$ and the melt crystallizes completely in the process of recalescence.\\
In 1900 Wilson already expressed his thoughts about the existence of a hypercooling limit \cite{Wilson1900}. In the 1960s Glicksman et al. described the hypercooled region \cite{GlSchae67} and mentioned the idea of a supercooled solid-liquid interface if $\Delta T_{\mathrm{hyp}}$ is exceeded \cite{Gl11}. Solidification from deep undercoolings is an up-to-date topic also in water research \cite{MaPaTaWöStWy12,XuPeSmKaKi16,BuBa16}. Only recently, Buttersack et al. reported about a new value for the hypercooling temperature of water \cite{BuWeBa18}.
In the late 1990s Wilde et al. \cite{WiGoeWi96} and Volkmann et al. \cite{VoWiWiHe98} mentioned $\Delta T_{\mathrm{hyp}}$ for completely miscible Pd-based alloys. The solidification velocity as a function of the undercooling was measured by Volkmann et al. with a short remark that passing the hypercooling limit has \textit{no influence} on $v(\Delta T)$ \cite{VoWiWiHe98}. However our findings raise doubts about this point. The scatter of our velocity data (y-axes of Fig. \ref{diagram_CuZr}, \ref{diagram_NiCuZr} and \ref{diagram_NiTi}) is at least two orders of magnitude smaller than the velocity measurements performed by Volkmann et al., which used photodiodes with electronical amplification.\\

In this work, we present experimental evidence for the influence of the hypercooling limit $\Delta T_{\mathrm{hyp}}$ on $v(\Delta T)$ for the binary intermetallic alloys NiTi and CuZr, as well as for the ternary compositions \ch{Cu_{0.6}Ni_{0.4}Zr}, \ch{Cu_{0.7}Ni_{0.3}Zr} and \ch{NiZr_{0.5}Ti_{0.5}}.

\section{Experimental}
\begin{figure}[tb]
	\begin{center}
		\includegraphics[width=8.6cm]{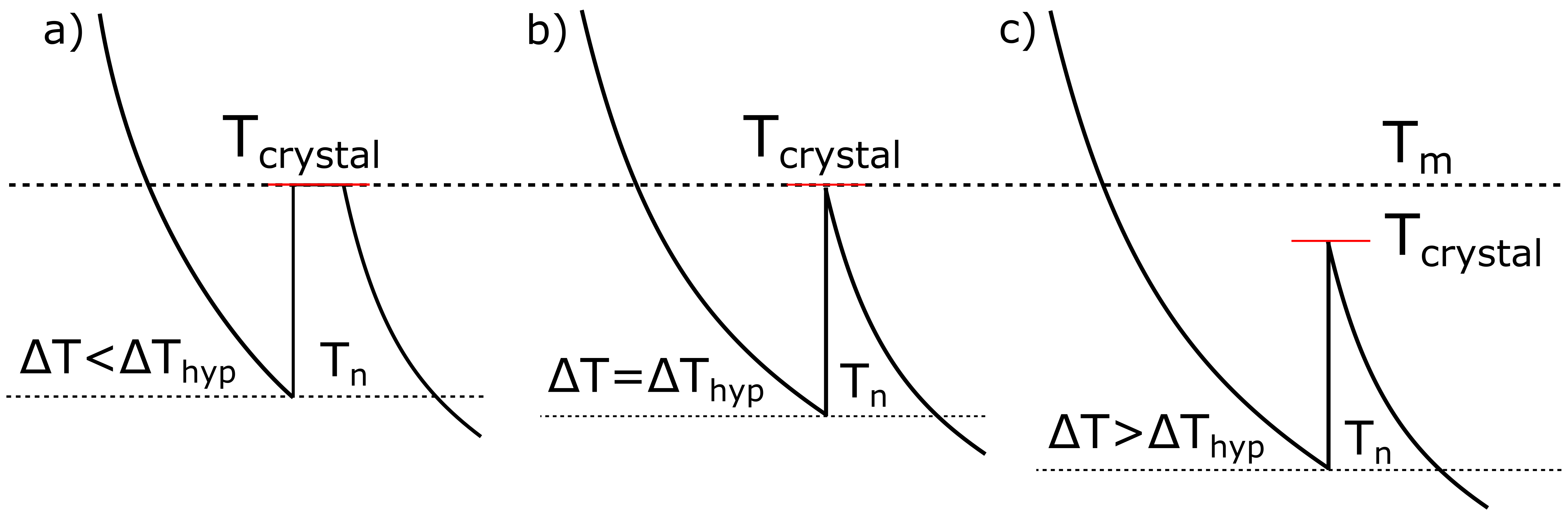}
	\end{center}
	\caption{\small{TTPs for three different undercoolings $T_\mathrm{n}$ illustrating the transition between $\Delta T < \Delta T_{\mathrm{hyp}}$ and $\Delta T > \Delta T_{\mathrm{hyp}}$ as it is observed for all systems investigated in this work. The time frame of the recalescence varies between $2\,\mathrm{ms}$ (NiTi) and $40\,\mathrm{ms}$ (Cu-Zr). a) At low and medium undercoolings $\Delta T < \Delta T_{\mathrm{hyp}}$ the latent heat that is released during solidification raises the sample temperature $T_\mathrm{crystal}$ up to the melting temperature $T_\mathrm{m}$. A part of the sample solidifies during the recalescence under non-equilibrium conditions. The rest solidifies in equilibrium in the melting plateau. b) If the released latent heat is just sufficient to raise $T_\mathrm{crystal}$ up to $T_\mathrm{m}$, the melting plateau vanishes. The specific undercooling temperature for this szenario corresponds to the hypercooling limit $\Delta T = \Delta T_{\mathrm{hyp}}$. c) For high undercoolings $\Delta T > \Delta T_{\mathrm{hyp}}$ the latent heat is not sufficient anymore to raise $T_\mathrm{crystal}$ up to $T_\mathrm{m}$ and $100\,\%$ of the sample solidifies under non-equilibrium conditions.}}
	\label{TTPs}
\end{figure}

High purity elements (Zr: $99.97\,\%$, Smart Elements; Ni: $99.995\,\%$, Alfa Aesar; Ti: $99.995\,\%$, Smart Elements; Cu: $99,999\,\%$, Smart Elements) are prepared and alloyed by arc-melting in a protective argon (6N) atmosphere yielding spherical reguli with a diameter of $4\,\mathrm{mm}$. The mass loss during arc-melting is found to be less than $0.2\,\mathrm{mg}$ for all alloys.  
A spherical sample is placed in an electrode system of an electrostatic levitator and is levitated and subsequently heated with a single infrared laser ($P = 75\,\mathrm{W}$, $\lambda = 808\,\mathrm{nm}$) under ultra-high vacuum conditions ($∼10^{-8}\,\mathrm{mbar}$). For a detailed description of the functionality of the ESL the reader is referred to reference \cite{DissMeister00}.
The temperature is measured contactless with a pyrometer (Impac IGA 120-TV, $1000\,\mathrm{Hz}$, $700-2300\,\mathrm{K}$, accuracy $\pm 5\,\mathrm{K}$).\\
A typical ESL cycle consists of the following temperature steps: The solid sample is heated up to $T_{\mathrm{m}}$. For congruently melting systems the temperature remains constant during the melting process yielding a plateau in the temperature-time-profile (TTP). If the sample is completely liquid, the temperature of the melt is further increased in order to evaporate possible contaminants. Subsequently, the infrared laser is switched off and the sample cools down only through radiation of heat. Due to the absence of crucible walls heterogeneous nucleation is being suppressed and the sample undercools well below $T_{\mathrm{m}}$ with a mean material specific cooling rate $\Delta T$/$\Delta t$. At some point i.e. at the nucleation temperature $T_{\mathrm{n}}$, the solidification sets in along with the release of latent heat, which increases the sample temperature. Due to the steep recalescence in the TTP, $T_{\mathrm{n}}$ can be pinpointed with high accuracy (see Fig. \ref{TTPs}).

With a high-speed camera (HSC) Photron Ultima APX 775k the crystallization of the liquid sample is recorded in order to ensure that only a single crystallization event occurs in every ESL cycle.
Table \ref{system data} shows the frame rates that were used for the different alloy systems due to their differing growth velocities.
For all cases a field of view of 512 x 320 pixel is chosen with a resolution of $\approx$ 30 $\mathrm{\mu}$m/pixel.

\section{Results and discussion}

\begin{figure}[tb]
	\begin{center}
		\includegraphics[width=8.6cm]{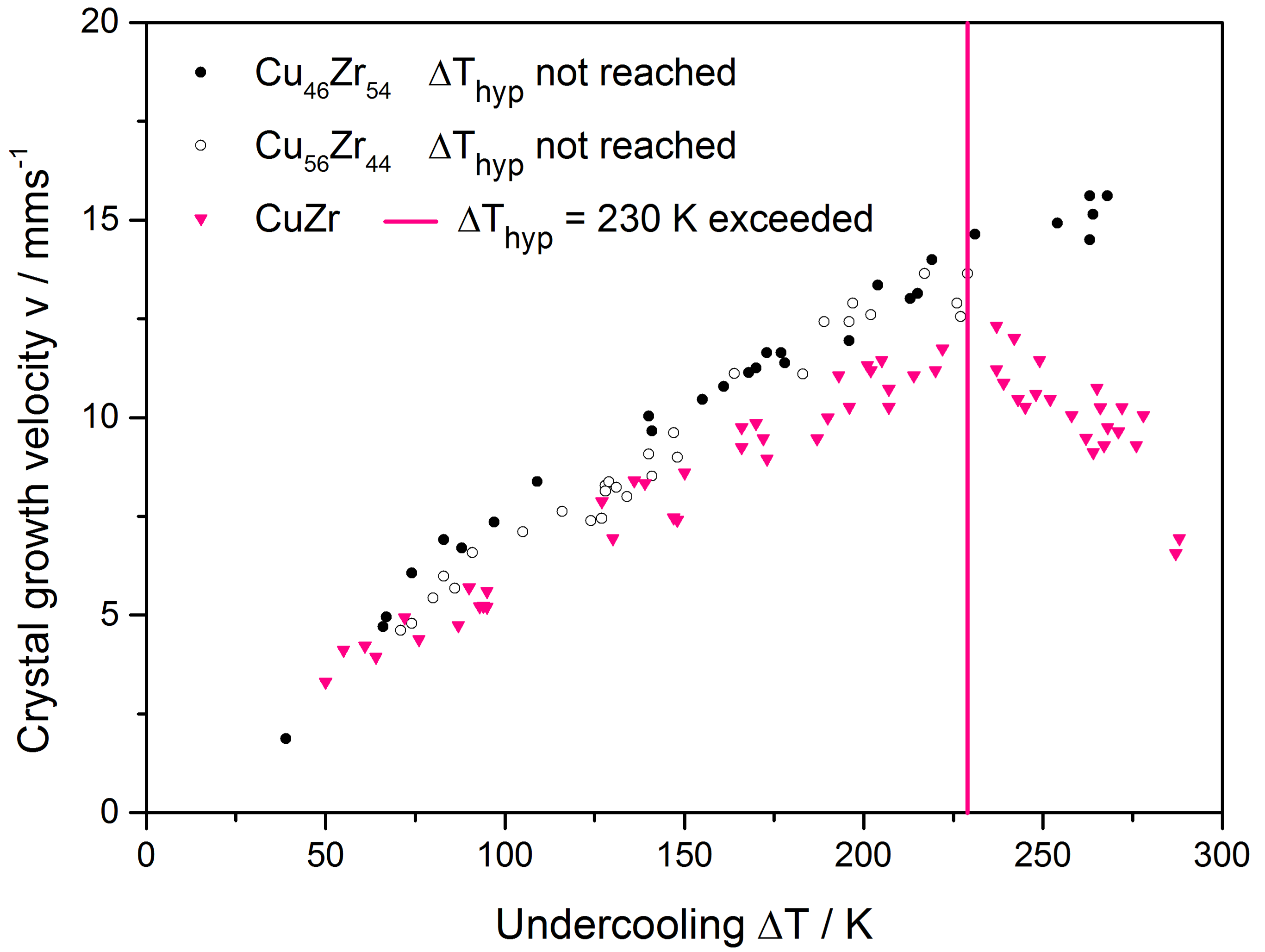}
	\end{center}
	\caption{\label{figproccham} \small{Crystal growth velocity $v$ as a function of the undercooling $\Delta T$ measured for intermetallic CuZr and the two adjacent eutectics \ch{Cu56Zr44} and \ch{Cu46Zr54}. While the $v(\Delta T)$ curve increases for both eutectics within the whole obtained undercooling regime, the curve for intermetallic CuZr increases until $\Delta T_{\mathrm{hyp}} = 230\,\mathrm{K}$ (vertical pink line) and then decreases with increasing undercooling.}}
	\label{diagram_CuZr}
\end{figure}

\begin{figure}[tb]
	\begin{center}
		\includegraphics[width=8.6cm]{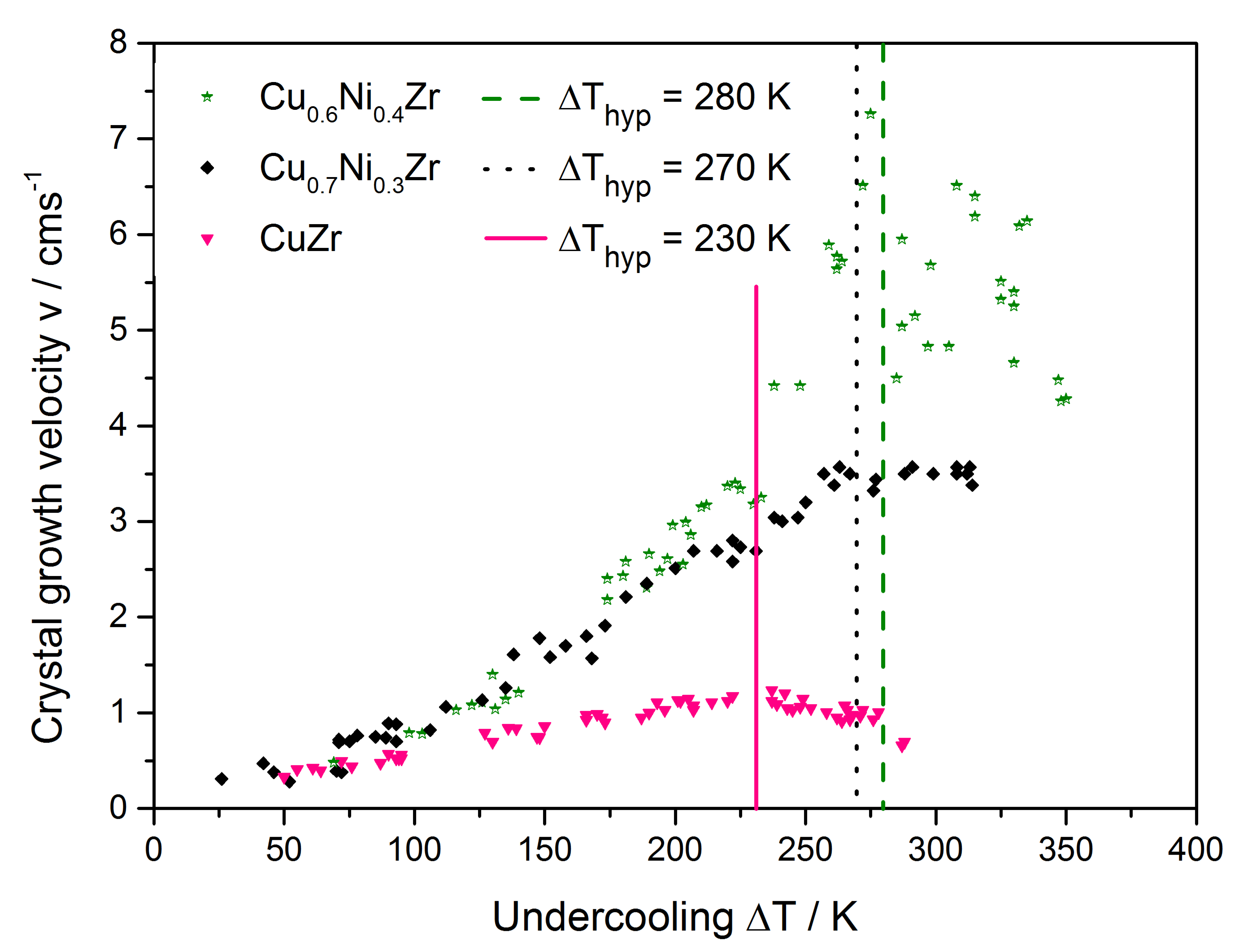}
	\end{center}
	\caption{\label{figproccham} \small{Crystal growth velocity $v$ as a function of the undercooling $\Delta T$ measured for the Zr-based ternary alloys \ch{Cu_{0.7}Ni_{0.3}Zr} and \ch{Cu_{0.6}Ni_{0.4}Zr}. One can see that, with increasing Ni at.\%, the velocity increases faster for $\Delta T < \Delta T_{\mathrm{hyp}}$ and $\Delta T_{\mathrm{hyp}}$ shifts to higher undercooling temperatures.  As observed for CuZr, the $v(\Delta T)$ relations for the ternary alloys exhibit a sudden change at $\Delta T_{\mathrm{hyp}}$.}}
	\label{diagram_NiCuZr}
\end{figure}

\begin{figure}[tb]
	\begin{center}
		\includegraphics[width=8.6cm]{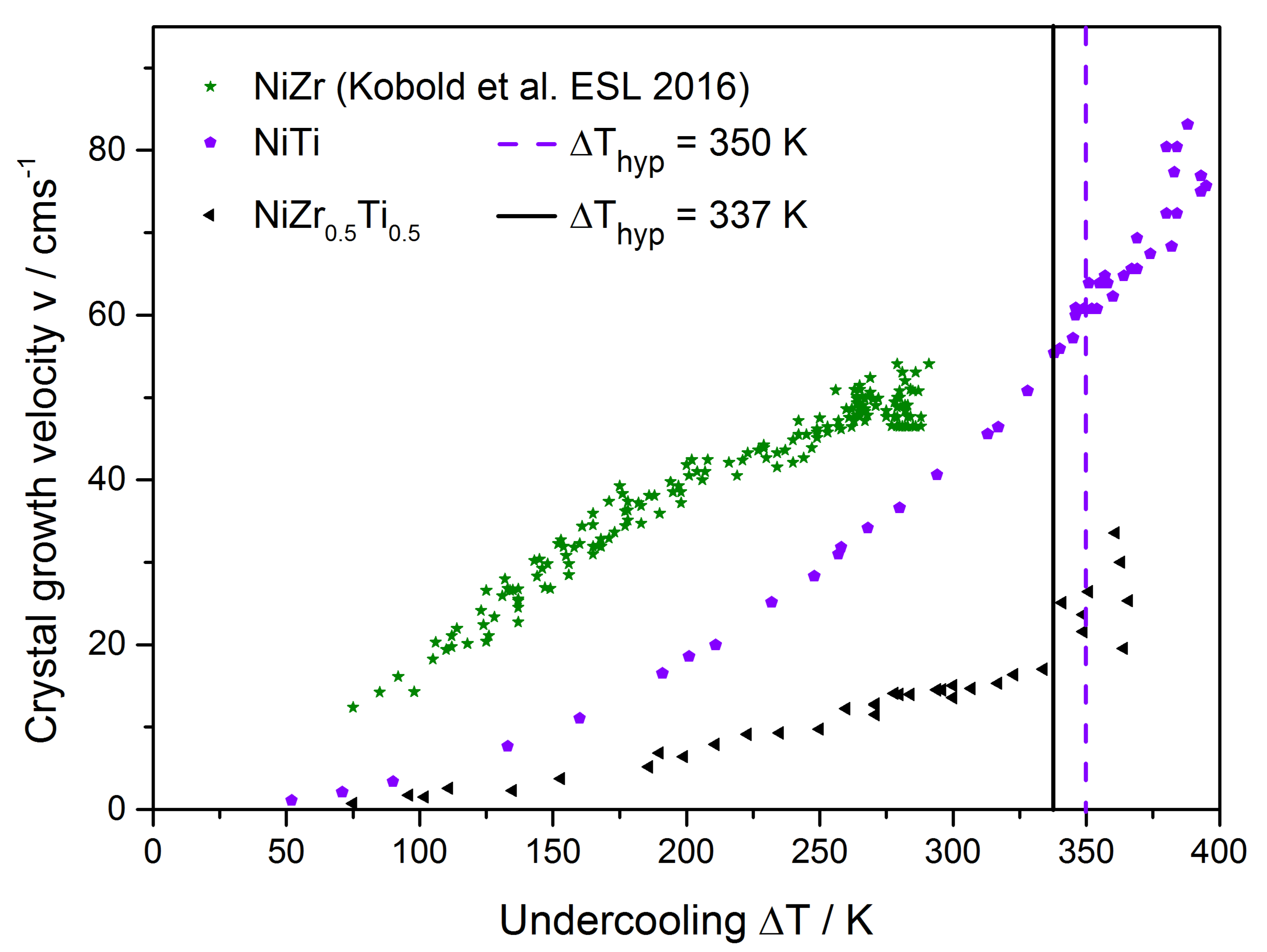}
	\end{center}
	\caption{\label{figproccham} \small{Crystal growth velocity $v$ as a function of the undercooling $\Delta T$ measured for intermetallic NiTi and the ternary alloy \ch{NiZr_{0.5}Ti_{0.5}}. NiTi exceeds the hypercooling limit significantly and shows a change in the $v(\Delta T)$ relation at $\Delta T_{\mathrm{hyp}}$. Therefore the Ni-based ternary alloy was predestined for exhibiting similar effects as observed for the (Cu$_{\mathrm{x}}$Ni$_{\mathrm{1-x}}$)Zr ($x= 0.7, 0.6$) alloys. Both systems show a similar behaviour for $\Delta T < \Delta T_{\mathrm{hyp}}$ as the NiZr system, whose $\Delta T_{\mathrm{hyp}}$ is not accessible \cite{KoKoHoHe18}.}}
	\label{diagram_NiTi}
\end{figure}

It was found that the solidification fronts are spherical for undercoolings $\Delta T > \Delta T_{\mathrm{hyp}}$ and can be approximated by a spherical shape for $\Delta T < \Delta T_{\mathrm{hyp}}$, respectively. Therefore the time between the first and last frame (FL-method) was used to determine the velocities. A spherical front is assumed to be formed by numerous dendrites growing in all directions i.e. it is most likely that a dendrite grows between the onset and end of solidification, which justifies using the FL-method to determine the dendrite growth velocity as a function of the undercooling. Accordingly, the $v(\Delta T)$ data was calculated with the ratio of the sample diameter $d_{\mathrm{s}}$ and the time $\Delta t$ between the first and last frame of the HSC recording i.e.  $v(\Delta T) = d_{\mathrm{s}}/\Delta t$.\\
Our $v(\Delta T)$ results for CuZr in Fig. \ref{diagram_CuZr} indicate that there is no continuous decrease in slope of the $v(\Delta T)$ curve as proposed in \cite{WaWaMaBiVoHeWaXuTiLi11,WaHeLi14}. In fact, we observe a kink in the curve of the solidification velocity at a specific undercooling temperature. No discrete change of the solidification front geometry is observed in the HSC recordings and hence this can be ruled out as being the cause of the observed deceleration. Remarkably, experiments we have carried out on a number of other systems corroborate our finding that the changes in the $v(\Delta T)$ relations can unambiguously be correlated with reaching the hypercooling limit (Fig. \ref{diagram_CuZr}, \ref{diagram_NiCuZr} and \ref{diagram_NiTi}). This observation is independent of the system's absolute solidification velocity as the absolute velocities of CuZr are in the range of $mm/s$ (Fig. \ref{diagram_CuZr}), while those of NiTi, (Cu$_{\mathrm{x}}$Ni$_{\mathrm{1-x}}$)Zr ($x= 0.6, 0.7$) and Ni(Zr$_{\mathrm{x}}$Ti$_{\mathrm{1-x}}) (x= 0.5)$ have values of $cm/s$ (Fig. \ref{diagram_NiCuZr} and \ref{diagram_NiTi}). For all $v(\Delta T)$ data points of a curve the frame rate of the HSC was kept constant i.e. the scatter becomes larger with increasing velocity. For every system the scatter at the highest velocity was calculated and is listed in Table \ref{system data}. Accordingly, these values represent the upper limit for the scatter of the velocity data of each system. The undercooling temperatures are precise to $\pm$ 5K.

In the case of CuZr (Fig. \ref{diagram_CuZr}), $\Delta T_{\mathrm{hyp}}$ is reached at an undercooling of $\Delta T = 230\,\mathrm{K}$. For $\Delta T < 230\,\mathrm{K}$ the $v(\Delta T)$ data increases. For $\Delta T > 230\,\mathrm{K}$ the slope of the $v(\Delta T)$ curve becomes negative and the velocity continues to decrease for higher undercoolings. The abrupt change in the $v(\Delta T)$ relation cannot be explained solely by $D(T)$ being dominant at high $\Delta T$. For the eutectics \ch{Cu56Zr44} and \ch{Cu46Zr54}, $\Delta T_{\mathrm{hyp}}$ was not reached within the obtained undercooling range. Coming from intermetallic CuZr, one can see that shifting the composition by $\pm$4 at. \% Zr engenders a drastic change of $\Delta T_{\mathrm{hyp}}$. Note that for both adjacent eutectic systems a similar behaviour is observed for $\Delta T < \Delta T_{\mathrm{hyp}}$.\\
Fig. \ref{diagram_NiCuZr} shows the measured $v(\Delta T)$ data for the ternary alloys \ch{Cu_{0.7}Ni_{0.3}Zr} and \ch{Cu_{0.6}Ni_{0.4}Zr}. For \ch{Cu_{0.7}Ni_{0.3}Zr} the measured hypercooling limit is $\Delta T_{\mathrm{hyp}} = 270\,\mathrm{K}$ and for \ch{Cu_{0.6}Ni_{0.4}Zr} it is $\Delta T_{\mathrm{hyp}} = 280\,\mathrm{K}$. As for CuZr, a significant change in the $v(\Delta T)$ relation can be pinpointed to these hypercooling limits. 
For \ch{Cu_{0.7}Ni_{0.3}Zr} we observe a power-law-like increase of $v(\Delta T)$ for $\Delta T < 270\,\mathrm{K}$ and a constant $v(\Delta T)$ for $\Delta T > 270\,\mathrm{K}$. For \ch{Cu_{0.6}Ni_{0.4}Zr} we also observe a power-law-like increase of $v(\Delta T)$ for $\Delta T < 280\,\mathrm{K}$ and a significant fluctuation of the $v(\Delta T)$ data for $\Delta T > 280\,\mathrm{K}$. Note that for these systems we observe a 4-fold soldification front in the HSC recordings for $\Delta T < \Delta T_{\mathrm{hyp}}$, which in this undercooling regime continuously evolves to a spherical shape with increasing undercooling and remains spherical for $\Delta T > \Delta T_{\mathrm{hyp}}$. Thus, the change of the $v(\Delta T)$ relation at $\Delta T_{\mathrm{hyp}}$ cannot be explained by a sudden change of the crystal morphology.\\
Fig. \ref{diagram_NiTi} shows the measured $v(\Delta T)$ data of intermetallic NiTi with a hypercooling limit of $\Delta T_{\mathrm{hyp}} = 350\,\mathrm{K}$ and the ternary alloy \ch{NiZr_{0.5}Ti_{0.5}} with $\Delta T_{\mathrm{hyp}} = 337\,\mathrm{K}$. Again, in both cases we can observe an influence on the $v(\Delta T)$ relation for $\Delta T > \Delta T_{\mathrm{hyp}}$.\\
Within a nucleation study Kobold et al. showed that for intermetallic NiZr $\Delta T_{\mathrm{max}} = 300\,\mathrm{K}$ is the highest undercooling achievable at a cooling rate of 34 $\mathrm{K/s}$ (vacuum of $10^{-7}\,\mathrm{mbar}$ in the electrostatic levitator) \cite{KoKoHoHe18}.
With $\Delta H_{\mathrm{f}}= 14.73\,\mathrm{kJ}/\mathrm{mol}$ and $c_{\mathrm{p}} = 42.6\,\mathrm{J}/\mathrm{(mol K)}$ one can calculate the hypercooling limit via equation \eqref{eq:Thyp} for congruently melting NiZr yielding an undercooling temperature of $\Delta T_{\mathrm{hyp}} = 346\,\mathrm{K}$ \cite{DissKobold16}. Therefore $\Delta T_{\mathrm{hyp}}$ was never reached for NiZr. Note that the kink in the velocity curve for NiZr at $\approx 175\,\mathrm{K}$ (Fig. \ref{diagram_NiTi}) does not correspond to the hypercooling limit of the system. The reason for this kink is a change in the crystal morphology, which will be discussed in a future publication.\\ 
We want to state again that the change in the $v(\Delta T)$ relations in Fig. \ref{diagram_CuZr}, \ref{diagram_NiCuZr} and \ref{diagram_NiTi} can clearly be linked to the undercooling temperature at which $100\,\%$ of the melt solidifies during recalescence, i.e. $\Delta T_{\mathrm{hyp}}$.

In the following we suggest one possible explanation (Fig. \ref{TTPs}) for our observations:
We define $T_{\mathrm{cn}} = T_{\mathrm{crystal}} - T_{\mathrm{n}}$, whereas $T_{\mathrm{crystal}}$ is the temperature of the growing crystal and $T_{\mathrm{n}}$ is the nucleation temperature.
$T_{\mathrm{cn}}$ becomes larger with increasing $\Delta T$, because $T_{\mathrm{crystal}}$ remains constant for $\Delta T < \Delta T_{\mathrm{hyp}}$  ($T_{\mathrm{crystal}} = T_{\mathrm{m}}$). For $\Delta T > \Delta T_{\mathrm{hyp}}$, $T_{\mathrm{crystal}}$ does not reach $T_{\mathrm{m}}$ anymore during recalescence, i.e. $T_{\mathrm{crystal}} < T_{\mathrm{m}}$. Therefore, compared to $\Delta T < \Delta T_{\mathrm{hyp}}$, $T_{\mathrm{cn}}$ should change differently for $\Delta T > \Delta T_{\mathrm{hyp}}$. This should effect the driving force for solidification, which usually increases linearly for $\Delta T < \Delta T_{\mathrm{hyp}}$. Consequently, exceeding $\Delta T_{\mathrm{hyp}}$ should have an effect on $v(\Delta T)$. If we consider NiTi (Fig. \ref{diagram_NiTi}), CuZr (Fig. \ref{diagram_CuZr}) and \ch{Cu_{0.7}Ni_{0.3}Zr} (Fig. \ref{diagram_NiCuZr}) the observed abrupt changes in the $v(\Delta T)$ relations at $\Delta T_{\mathrm{hyp}}$ could be explained. For $\Delta T > \Delta T_{\mathrm{hyp}}$ the different cases (NiTi: increase, CuZr: decrease, \ch{Cu_{0.7}Ni_{0.3}Zr}: constant) cannot be unequivocally connected to how $T_{\mathrm{cn}}$ changes, but one has to consider also other (kinetic) factors. However, for \ch{Cu_{0.6}Ni_{0.4}Zr} (also Fig. \ref{diagram_NiCuZr}), $T_{\mathrm{cn}}$ should be irregular due to the large scatter in $v(\Delta T)$ for $\Delta T > \Delta T_{\mathrm{hyp}}$.

\section{Outlook}
We are convinced that the discovery of the effect of the hypercooling limit on the crystal growth velocity will influence the understanding of solidification processes under non-equilibrium conditions. Fields, in which non-equilibrium processes play a dominant role, will benefit from an improvement of existing growth models:

In industrial manufacturing processes for instance, in which the material is directly crystallized from the undercooled melt, it is crucial to understand the solidification behaviour to be able to predict the microstructure of the obtained product more precisely. For example, Flow Induced Crystallization (FIC) is a material production process used daily in polymer industry \cite{WaJuYaMaLiCuYaChHuLi16}.

In the field of climate research, precipitation models are essential in understanding the water cycle of the earth. Non-equilibrium solidification governs, for instance, the formation of hailstones or ice clouds \cite{Pr97}.

In astrogeophysics, the structure and formation of meteorites provides insight into the genesis of our solar system. Remarkably, Nagashima et al. reproduced chondrules from levitated, hypercooled melts of forsterite ($\ch{Mg2SiO4}$) in the laboratory \cite{NaTsSaKoDo06}. Primitive meteorites are remnants of the solar nebula, containing intermixed rock and metal flakes. Chondrules are small spherical silicate inclusions in these meteorites, which were formed under ambient conditions of our early solar system.
Thus, understanding the formation conditions of chondrules gives decisive information about the conditions, which prevailed in the early period of our solar system.

\section{Summary and Conclusion}
We have investigated the crystal growth velocity as a function of the undercooling temperature for the glass forming alloys NiTi and Cu--Zr, as well as for the Zr-based ternary alloys (Cu$_{\mathrm{x}}$Ni$_{\mathrm{1-x}}$)Zr ($x= 0.6, 0.7$) and the Ni-based ternary alloy Ni(Zr$_{\mathrm{x}}$Ti$_{\mathrm{1-x}}) (x= 0.5)$ with an improved experimental setup, which allowed us to significantly reduce the scattering of the data and make small effects on the $v(\Delta T)$ curves visible. In contrast to previous publications on CuZr \cite{WaWaMaBiVoHeWaXuTiLi11,WaHeLi14} we do \textit{not} observe the $v(\Delta T)$ behaviour as described in these publications. For all investigated systems we observe a crystal growth velocity which, for $\Delta T < \Delta T_{\mathrm{hyp}}$, increases with a power law with increasing undercooling temperature and a decreasing (CuZr), a constant (\ch{Cu_{0.7}Ni_{0.3}Zr}) and scattering (\ch{Cu_{0.6}Ni_{0.4}Zr}, \ch{NiZr_{0.5}Ti_{0.5}}) behaviour for $\Delta T > \Delta T_{\mathrm{hyp}}$.\\
Since these behaviours show a clear connection to $\Delta T_{\mathrm{hyp}}$, we do \textit{not} assume that the underlying effect is caused by glass forming properties (CuZr, (Cu$_{\mathrm{x}}$Ni$_{\mathrm{1-x}}$)Zr ($x= 0.7, 0.6$)) in the obtained undercooling regimes. The observed effect seems to be fundamental for alloy systems which exceed the hypercooling limit. To our understanding, this effect has not yet been addressed in any scientific work, even though several articles on hypercooling in combination with $v(\Delta T)$ analysis have been published \cite{WiGoeWi96,VoWiWiHe98,GlSchae67}. With our results we could shed new light on the existing paradigm for the reason of the deceleration effect observed in the solidification velocity with increasing undercooling temperature in glass forming systems.  In general, this has far-reaching ramifications in various fields as diverse as materials science, climate research and astrogeophysics.

\appendix
\section{Experimental}
Fig. \ref{HSC_frames} shows two frames from a HSC recording of the solidification process of CuZr.
A small mirror, which is installed behind the levitating sample and in the line of sight of the HSC, makes it possible to pinpoint the onset and end of solidification to a single frame yielding a significant enhancement of the accuracy of the measured $v(\Delta T)$ data. In previous works, where $v(\Delta T)$ was determined with a HSC, only $50\,\%$ of the sample surface were recorded. Note that in many cases the onset or end of solidification takes place at the part of the sample that is not visible. Therefore the remaining $50\,\%$ were linearly extrapolated \cite{AsReHe06,WaWaMaBiVoHeWaXuTiLi11}. However, for most intermetallic alloys the solidification front does not propagate with constant velocity over the sample surface, although the growth velocity of the crystal growing into the undercooled melt is assumed to be constant. This is due to the intersection between the spherical sample surface and the often complex geometry of the growing crystal \cite{DissKobold16}. Therefore an extrapolation to the remaining $50\,\%$ often leads to a large scatter, especially if the crystal geometry is unknown (as is the case for all investigated alloys in this work). The scatter can be significantly reduced in our setup, where we use a mirror in combination with a high frame rate.

\section{Results and discussion}

With Planck`s law one can calculate the specific heat capacity $c_{\mathrm{p}}$($T_{\mathrm{L}})$ at the liquidus temperature as described by Wessels et al. \cite{WeGaSaHyCaRoKrGoRoLeMoKe11} via
\begin{equation*}
	c_{\mathrm{p}}(T_{\mathrm{L}})= \frac{\- \sigma_B \cdot A \cdot \epsilon_T (T^4-{T_0}^4)}{m \cdot \frac{\mathrm{d}T}{\mathrm{d}t}}
\end{equation*}
with $\sigma_B$: Stefan-Boltzmann constant, A: sample surface area, $\epsilon_T$: hemispherical total emissivity, T: temperature of the melt taken from the TTP and $T_0$: ambient temperature. Since the temperature dependency of $\epsilon_T$ is unknown, a material specific constant value ($\epsilon_T= 0.28$) is used to calculate $c_{\mathrm{p}}$. The cooling rate $\mathrm{d}T/\mathrm{d}t$ is determined from the part of the TTP, where the sample cools down only by radiation of heat.
With $\Delta H_{\mathrm{f}}$ \cite{WaWaMaBiVoHeWaXuTiLi11, DissKrause02} and $c_{\mathrm{p}}$ (this work, see Fig. 6 and 7) the hypercooling limits for the investigated Cu-Zr alloys were calculated (see Table \ref{system data}). Note that for CuZr the theoretically determined $\Delta T_{\mathrm{hyp, calc}}$ and the experimentally determined $\Delta T_{\mathrm{hyp, exp}}$ match perfectly. Note also that the $\Delta T_{\mathrm{hyp, calc}}$ values for the eutectic compositions \ch{Cu56Zr44} and \ch{Cu46Zr54} are higher than the maximum undercoolings $\Delta T_{\mathrm{max}}$ reached in our experiments.

\begin{table*}[htb]
	\caption{Enthalpies of fusion $\Delta H_{\mathrm{f}}$, specific heat capacities at constant pressure $c_{\mathrm{p}}$, calculated hypercooling limits $\Delta T_{\mathrm{hyp, calc}}$ (via eq. (\ref{eq:Thyp})), experimentally found hypercooling limits $\Delta T_{\mathrm{hyp, exp}}$ (extracted from the TTPs), maximum achieved undercooling temperatures $\Delta T_{\mathrm{max}}$, the errors for the highest measured velocities $\Delta v$ and the used frame rates for the HSC in frames per second (fps)}
	\setlength{\tabcolsep}{13pt} % General space between cols (6pt standard)
	\renewcommand{\arraystretch}{1.25} % General space between rows (1 standard)
	\begin{tabular}{l|ccccccl}\toprule
		\textbf{System}	&\textbf{$\Delta H_{\mathrm{f}}$}	&\textbf{$c_{\mathrm{p}}$}	&\textbf{$\Delta T_{\mathrm{hyp, calc}}$}	&\textbf{$\Delta T_{\mathrm{hyp, exp}}$} &\textbf{$\Delta T_{\mathrm{max}}$}	&\textbf{$\Delta v$}	&fps	\\
		& kJ/mol & J/(mol K) & K & K & K & cm/s & 1/s \\\midrule
		\hline
		\ch{Cu56Zr44} 			&	13.7$^a$	&	40.41		& 339	& -		&	230		& $\pm$ 0.2		& 500 		\\
		\ch{CuZr} 				&	9.219$^b$	&	39.72		& 232	& 230	&	280		& $\pm$ 0.2		& 500	  	\\
		\ch{Cu46Zr54} 			&	12.5$^a$	&	39.28		& 318	& -		&	270		& $\pm$ 0.2		& 500  		\\
		\ch{Cu_{0.7}Ni_{0.3}Zr}	&	11.1$^d$	&	41.22$^d$	& -		& 270	&	320		& $\pm$ 0.5		& 500 		\\
		\ch{Cu_{0.6}Ni_{0.4}Zr}	&	10.6$^d$	&	37.83$^d$	& -		& 280	&	350		& $\pm$ 0.1		& 2000 		\\
		\ch{NiZr}				&	14.73$^c$	&	42.6$^c$	& 346	& -		&	300		& $\pm$ 0.5		& 10000 	\\
		\ch{NiZr_{0.5}Ti_{0.5}}	&	15.6$^d$	&	46.3$^d$	& -		& 337	&	360		& $\pm$ 0.5		& 6000 		\\
		\ch{NiTi}				&	12.5$^d$	&	35.66$^d$	& -		& 350	&	390		& $\pm$ 0.8		& 20000 	\\	\bottomrule
	\end{tabular}
	\label{system data}
	\begin{tablenotes}\footnotesize 
		\item a:\cite{DissKrause02}; b:\cite{WaWaMaBiVoHeWaXuTiLi11}; c:\cite{DissKobold16}; all other values are calculated or measured in this work
		\item d: The $c_{\mathrm{p}}$ values were calculated from the cooling curves of the experiment and averaged. The $\Delta H_{\mathrm{f}}$ values were then determined via $\Delta T_{\mathrm{hyp, exp}}$. The errors for $c_{\mathrm{p}}$ are within $\pm\,0.3\,\mathrm{J}/\,\mathrm{(molK)}$ and for $\Delta H_{\mathrm{f}}$ within $\pm\,0.1\,\mathrm{kJ}/\,\mathrm{mol}$ (for \ch{Cu_{0.6}Ni_{0.4}Zr}: $\pm\,2.41\,\mathrm{J}/\,\mathrm{(molK)}$ and $\pm\,1\,\mathrm{kJ}/\,\mathrm{mol}$).
	\end{tablenotes}
\end{table*}

\newpage
\begin{figure}[h]
	\begin{center}
		\includegraphics[width=8.6cm]{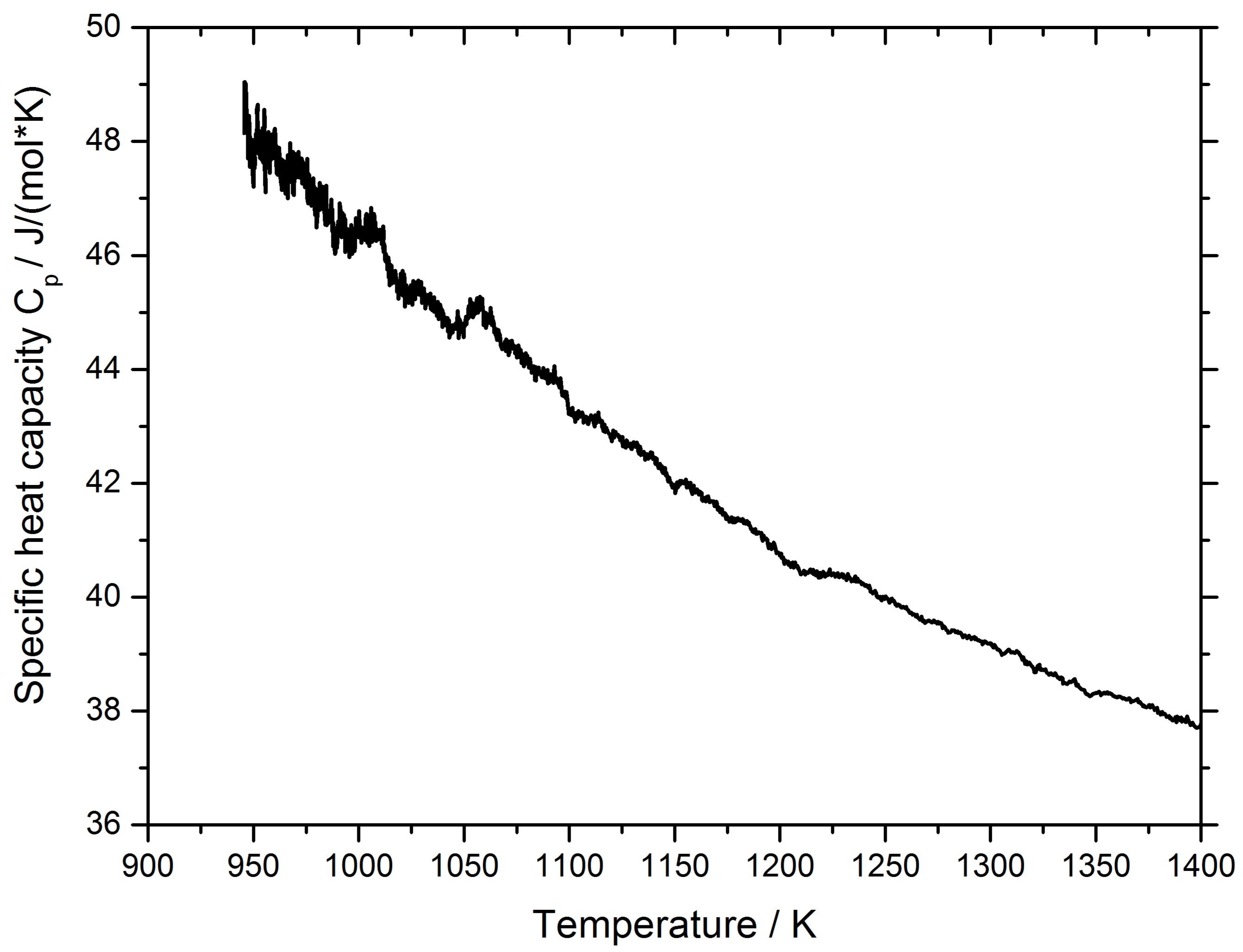}
	\end{center}
	\caption{\label{figproccham} \small{Specific heat capacity $c_\mathrm{p}$ as a function of temperature as calculated from the TTP of CuZr.}}
	\label{cpps}
\end{figure}

\begin{figure}[h]
	\begin{center}
		\includegraphics[width=8.6cm]{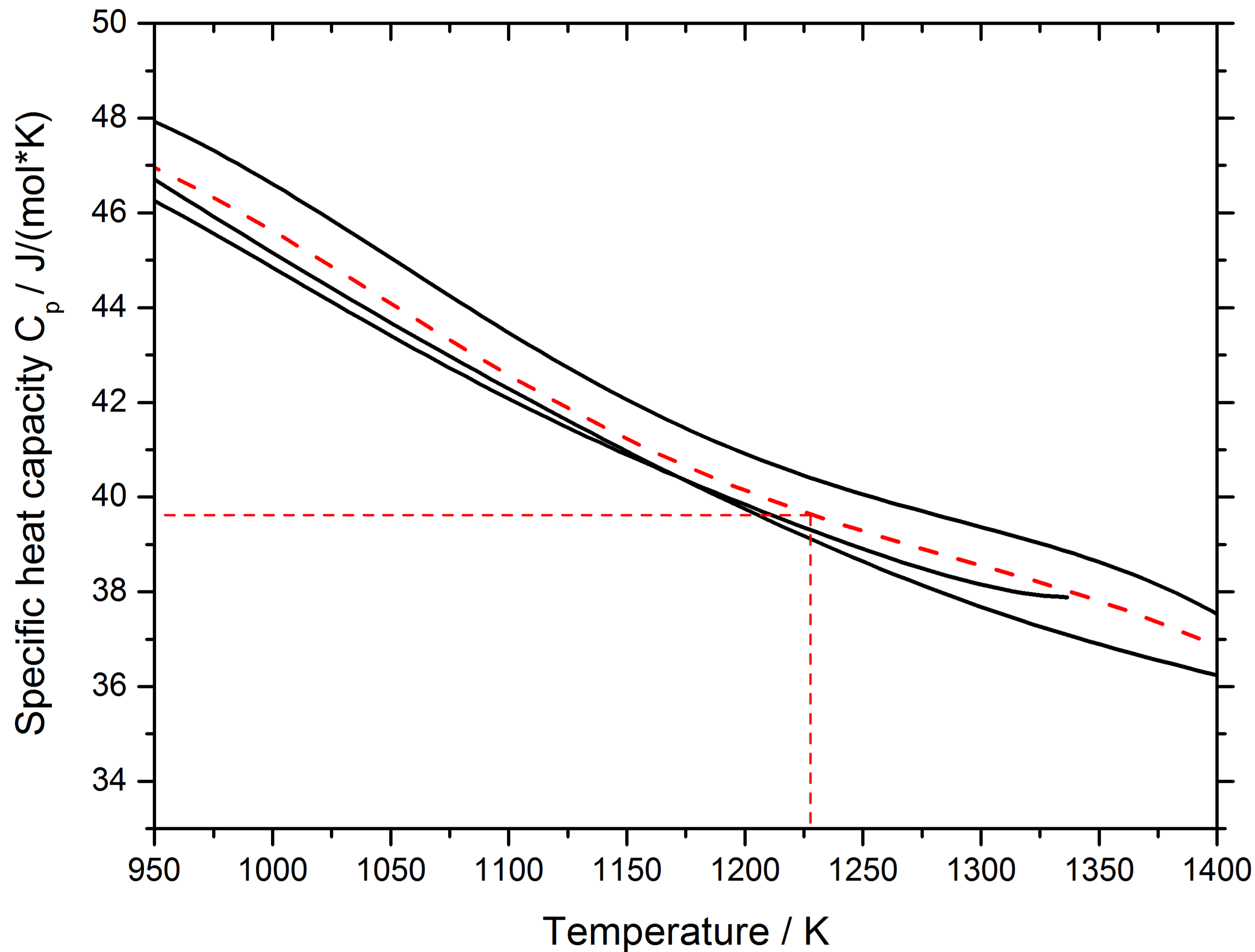}
	\end{center}
	\caption{\label{figproccham} \small{Every black curve represents the average of three single $c_\mathrm{p}$ measurements for CuZr as seen in Fig. \ref{cpps}. The dashed red curve is an average of the three black curves in the diagram. The specific heat capacity $c_{\mathrm{p}}= 39.72\,\mathrm{J} / \mathrm{(mol K)}$ is marked at the liquidus temperature $T_{\mathrm{L}}=1226\,\mathrm{K}$ of CuZr.}}
	\label{cpp}
\end{figure}

\pagebreak

\end{document}